*Review*

# A review of urban resilience frameworks: Transferring knowledge to enhance pandemic resilience

**Yue Sun[1,*], Ryan Weightman[2], Anye Shi[1], Timur Dogan[1] and Samitha Samaranayake[1]**

[1] Cornell University, Ithaca, NY 14853, USA
[2] Center for Computational and Integrative Biology, Rutgers University-Camden, Camden, NJ 08102, USA

* **Correspondence:** Email: ys954@cornell.edu.

**Abstract:** Urbanization is rapidly increasing, with urban populations expected to grow significantly by 2050, particularly in developing regions. This expansion brings challenges related to chronic stresses and acute shocks, such as the COVID-19 pandemic, which has underscored the critical role of urban form in a city's capacity to manage public health crises. Researchers are examining how built form, density, and neighborhood structure shape pandemic outcomes. This work adds to the literature and, rather than offering another empirical case study or city-specific design checklist, it bridges the general resilience/morphometrics frameworks with pandemic-specific discussions. We seek to transfer the knowledge of prevalent resilience theories into the context of a pandemic to better understand the relationship between pandemic resilience and urban morphology. In this work, we emphasized the connection between theoretical ideas with quantitative frameworks, offering a cohesive analysis that traces how concepts such system assets, attributes, and performances can be operationalized through morphometric indicators. We synthesize general urban resilience frameworks and morphometric approaches to explicitly map how their concepts and indicators can be adapted to the assessment of pandemic resilience while engaging with the emerging literature on pandemic urbanism and extended urbanization. Our goal is not only to enhance the understanding of urban resilience but also to offer a multi-scalar conceptual map for future empirical work.

## 1. Introduction

### *1.1. Resilience as a powerful lens for urban planning and design*

The United Nations projects that by 2050, urban populations will grow by 2.5 billion, especially in Africa and Asia [1]. This expansion will significantly impact cities' physical infrastructure, including buildings, street networks, and public service systems. Cities face challenges like housing shortages, air pollution, inefficient transport systems, and exposure to extreme heat [2,3]. Because the built environment is capital-intensive and slow to change, these stresses can become locked in for decades. The COVID-19 pandemic further underscored the importance of urban morphology and urban planning in managing public health crises, as researchers have shown how density, land-use patterns, access to green space, and housing conditions relate to infection risks and quality of life [4–8]. Given these challenges, the concept of resilience provides a valuable perspective for understanding and addressing ongoing changes in the cities of our world [9].

We have four major objectives. First, we identify the principal conceptual interconnections between urban resilience and urban form. Second, we review frameworks, with an emphasis on quantitative approaches, that operationalize general urban resilience through variables and metrics grounded in urban morphology. Third, by putting these frameworks into a broader perspective of general urban resilience, we aim to develop a more systematic understanding of how urban form and design factors affect a city's ability to withstand and recover from disruptions. Fourth, we explore how this knowledge can be transferred to deal with situations such as the COVID-19 pandemic. In this review, we highlight gaps in frameworks, provide insights into physical properties that are essential for mitigating the impacts of pandemics, and suggest areas for improvement, thereby contributing to the development of more adaptive frameworks for assessing urban resilience in the face of future crises.

Guided by these objectives, we address the following three research questions: i) How have urban resilience frameworks conceptualized the role of urban form and urban morphology in system resilience? ii) What are quantitative, morphometric indicators that are commonly used to operationalize resilience attributes, and how do they relate to urban form elements? iii) How can these frameworks and indicators be adapted or extended to assess pandemic resilience in urban environments? Given the conceptual and synthetic nature of this article, we do not test statistical hypotheses. Instead, we work with the following guiding hypothesis: General urban resilience frameworks, developed largely without explicit reference to pandemics, embed morphometric attributes (such as diversity, connectivity, modularity, redundancy, efficiency, and density) that are theoretically relevant to pandemic resilience and can be explicitly transferred to pandemic-resilience assessments.

### *1.2. Contextualize pandemic resilience into broader urban resilience frameworks*

Despite the growing interdisciplinary adoption of urban resilience, research that explicitly links urban form and urban morphology to pandemic resilience is emerging. A number of reviewers and empirical researchers have examined how compactness, urban design, housing conditions, and neighborhood structure influence infection risks, quality of life, and vulnerability during COVID-19 [4–8,10,11]. These studies are dispersed across disciplines and focus on specific indicators (such as density, crowding, or access to green space) or on case-study cities. Building on these contributions and on emerging debates around "pandemic urbanism" and extended urbanization [12–15], our review seeks to contribute a synthesis of how general urban resilience frameworks and morphometric indicators can be translated into the pandemic context and used to structure pandemic-resilience assessment in cities.

As a subset of general resilience, urban resilience to pandemics focuses on an urban area's ability to manage and mitigate the impacts of infectious disease outbreaks. Since the principles underpinning pandemic resilience are closely tied to those of general resilience, and urban systems are often intricately interconnected, it is crucial to move beyond addressing individual hazards in isolation. Such an approach risks producing assessments and prioritizations that fail to account for interconnected impacts. Hence, we integrate pandemic resilience into broader urban resilience frameworks and examine the relationship between general resilience and urban form to pinpoint crucial aspects of morphological and systemic features specific to public health events and to develop guidance on effective spatial planning strategies that bolster adaptive capacity.

In this article, we use "urban resilience" in line with Meerow et al. [9] as the capacity of urban systems, communities, individuals, organizations, and businesses to survive, adapt, and grow in the face of chronic stresses and acute shocks. "Pandemic resilience" denotes the subset of this capacity related to infectious disease threats, including the ability to prevent, absorb, and recover from epidemic shocks while maintaining essential functions [4,15]. We use "urban form" to refer to the spatial configuration of streets, blocks, plots, buildings, and open space, and "urban morphology" to emphasize the multi-scalar patterns and evolution of this configuration [16,17].

To comprehend the drivers and interactions among natural, human-built, and social systems in urban areas and their effects on sustainability outcomes at various scales, researchers have identified two crucial areas of knowledge: Resilience theory and urban morphology [16,18]. Emerging from the field of ecology in the 1970s to describe the capacity of a system to maintain or recover functionality in the event of disruption or disturbance, the term resilience has been increasingly applied across a growing number of fields, such as critical infrastructures [19], economy [20], and supply chain [21]. However, consensus on its definition remains elusive due to the varying contexts where it is applied. Resilience could be applicable in the context of cities because cities are complex systems adapting to changing circumstances [22]. The dynamic nature of complex urban systems is highly improbable for a full post-disturbance return to a previous state [23]. Thus, a system's capability to maintain key functions, while not necessarily returning to a prior state, is of vital importance.

Sharing similar objectives, urban morphology, as the result of strategic manipulation in urban planning and design, bears the function to steer socio-economic and environmental changes toward targeted outcomes. Its patterns, which detail the structure, formation, and transformation of the environment, are crucial for understanding how components of a city work together. They are also important in addressing human needs and accommodating cultural practices. Therefore, to make "resilient" thinking more practically applicable to urban planning and design, the concepts of resilience are ideally expressed through specific language that planners and designers can utilize (i.e., urban morphology) [24,25].

The marriage of the two fields highlights the notion of a resilient city, which gains relevance when chronic stresses or abrupt shocks pose a threat of widespread disruption. The combination can serve as a bridge between disaster risk reduction and conventional urban planning and design, shifting focus from conventional disaster risk management, which tends to be hazard-specific, to a more inclusive approach that anticipates various unpredictable events [3]. This is what Miller et al. [26] claim is the transition from "specified" to "general" resilience. In the work, they point out that focusing excessively on specified resilience can undermine overall system flexibility, diversity, and the ability to respond to unexpected threats. In this context, solely emphasizing the interconnection between resilience to a pandemic, and urban form might ignore implications and unforeseen costs in other areas. Thus, a comprehensive view toward urban resilience is needed that not only recognizes unique pandemic impacts, but also maintains event-agnostic ways to embrace adaptability to inherently unstable equilibria of urban systems.

### *1.3. Structure of the review*

The rest of this paper is divided into several key sections (Figure 1). In Section 2.1, we delve into the significance of urban form to a city's resilience. In Section 2.2, we explore their multifaceted and interdependent relationship, illustrating how urban form functions as a complex system with various interrelated layers that affect the city's response capabilities to disruptions. In Section 2.3, we introduce key theoretical frameworks that link resilience with urban form. Concepts such as Complex Adaptive Systems (CAS), the Adaptive Cycle Model, and Panarchy Theory are discussed to provide a deeper understanding of how resilience operates within urban contexts and how it can be applied to analyze urban form. In Section 2.4, we focus on quantitative frameworks used to assess urban resilience. In Section 3.1, we discuss gaps in frameworks. Moreover, we emphasize the need for more adaptive and integrative approaches that can better account for the complexities of urban systems in the face of pandemics. In Section 3.2, we suggest ways in which frameworks can be enhanced to provide more comprehensive and adaptive assessments of pandemic resilience. In Section 3.3, we discuss limitations and scope of the work.

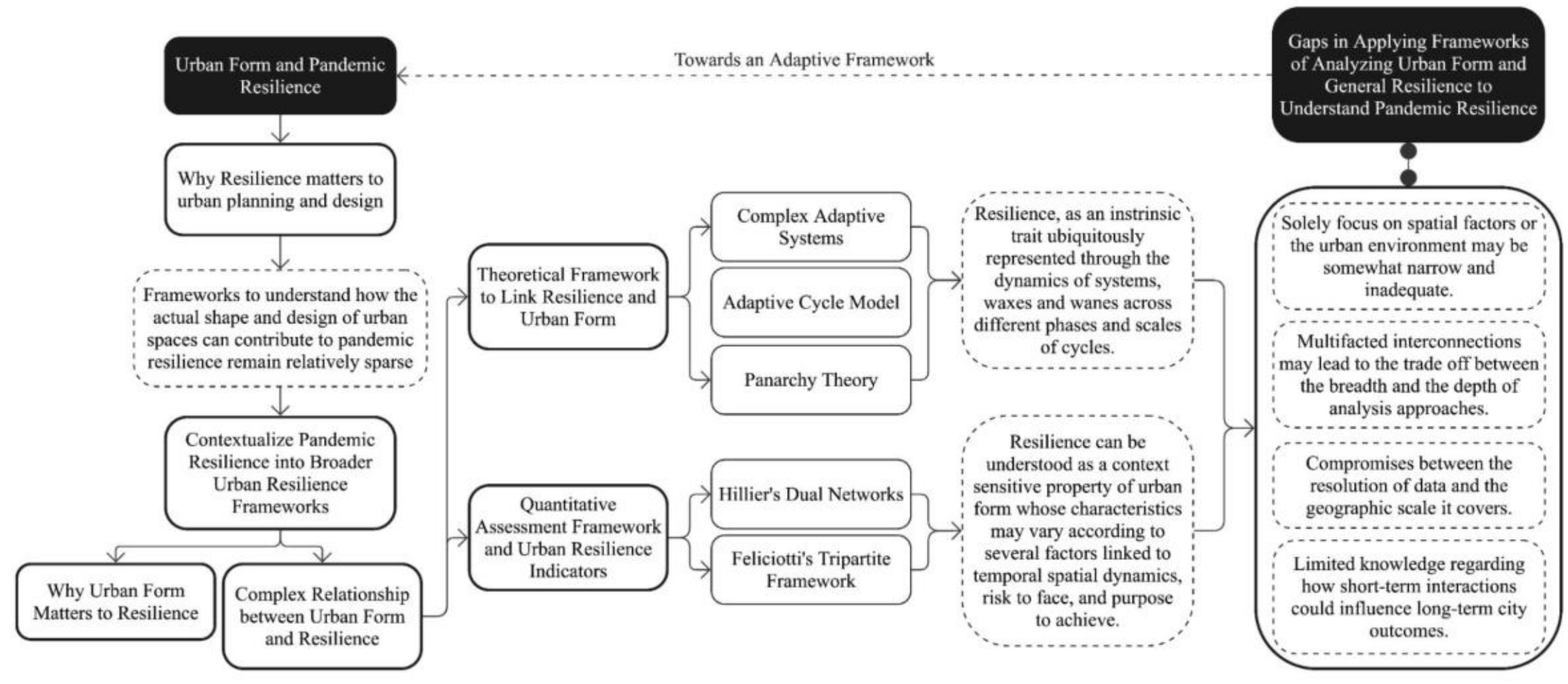

**Figure 1.** Structure of the review.

## 2. Literature review

Our work is designed as a narrative, critical review that synthesizes conceptual and methodological frameworks at the intersection of urban resilience, urban morphology, and pandemic resilience, rather than as a systematic or bibliometric review in the PRISMA sense. We identify relevant studies through iterative searches in Web of Science and Scopus, complemented by Google Scholar, using combinations of keywords such as ("urban resilience" OR "resilient city") AND ("urban form" OR "urban morphology" OR "urban design") AND ("framework" OR "indicator" OR "metric") and, for the pandemic-focused literature, ("COVID-19" OR "pandemic") AND ("urban form" OR "compact city" OR "density"). We then use backward and forward citation tracking to refine the corpus and to identify widely cited frameworks and exemplary applications. Given our focus on frameworks and morphometric indicators, we concentrate on peer-reviewed journal articles and books that i) explicitly conceptualize urban resilience and include urban form, ii) propose quantitative morphometric indicators for resilience assessment, or iii) analyze pandemic resilience with explicit reference to urban form. In keeping with the aims of a critical, theory-building review, we do not attempt to exhaustively enumerate all publications on these topics or provide formal bibliometric statistics. Instead, we use selected influential contributions to build an integrated conceptual map.

### *2.1. Why urban form matters to resilience*

Urban form is as a subset of urban infrastructure and form that is one of four subsystems identified by [9,27], as shown in Table 1. It includes key elements such as streets, plots, blocks, and open spaces. While these physical components may appear rigid compared with resilience characteristics which highlight the adaptation to the state of constant changing non-equilibrium [18], their impacts for achieving urban resilience may be substantial. Urban physical structures are considered the raw material of urban planning and design [28]; they shape the flow of people and goods through a city [29]. During the outbreak of the COVID-19 pandemic, houses were shown to be crucial sanctuaries for individuals against the coronavirus, with those lacking homes being the most vulnerable [30]. Interruptions in vital infrastructure can disrupt routine flows and result in extensive damage. For instance, the virus has broadly transformed the landscape of critical infrastructure. Batty [25] expressed concerns over potentially structural shift of workplace from office to home. This might cause a dramatic dent in the public transportation sector and commercial real estate sector. A survey carried out by Galbusera et al. [31], involving key stakeholders in infrastructure operation, reveals that, aside from a few key sectors such as the health sector, information and communication technology, and water utilities, most industries within the critical infrastructure field experienced negative impacts in terms of demand, supply, operation, and profitability.

On the other hand, urban form, due to its long-lasting nature, can lock cities into positive or negative development trajectories [9]. Cities lacking resilient design risk becoming entrenched in unsustainable paths, negatively impacting long-term livability. For instance, car-centric cities faced significant challenges during the pandemic, with increased reliance on private vehicles leading to higher traffic and air pollution as public transport usage declined due to fears of virus transmission [32]. This shift exposed the fragility of car-dependent systems. In contrast, cities like Paris and Milan, which had invested in sustainable urban forms, adapted better by reducing car dependency and promoting active transportation, such as bike lanes and pedestrian spaces. These measures supported social distancing, improved air quality, and reduced carbon emissions. Access to green spaces and walkable neighborhoods also proved essential during lockdowns, contributing to better physical activity and mental health [33]. This highlights the role of integrating micro-scale factors into urban design to enhance resilience and overall livability, ensuring that cities are better equipped to handle future crises.

**Table 1.** Conceptual categorization of urban systems.

| Subsystems | Elements | Description | System dynamics |
|---|---|---|---|
| Governance networks | States, Industry, Consumers, Labor, NGOs | Governance networks involve diverse actors and institutions, including government levels, NGOs, and businesses, shaping urban systems through decision-making. | All the elements within urban systems are multi-scalar, networked, and often strongly coupled, featuring interconnections both within and between four complex and adaptive subsystems that interact across various spatial and temporal scales. |
| Networked material and energy flows | Waste, Energy, Materials, Water, Food, Consumer Goods | This layer encompasses the myriad materials and energy flows produced or consumed within urban systems, like water, energy, and waste. | |
| Urban infrastructure and form | Buildings, Utilities, Ecological Greenspace, Transportation | Focuses on the built environment, including buildings, transportation networks, and utilities, along with urban green spaces and parks. | |
| Socio-economic dynamics | Demographics, Mobility, Equity and Justice, Public Health, Capital, Education | It covers aspects like monetary capital, demographics, equity and justice, which influence other subsystems as well as the livelihoods and capacities of urban citizens. | |

### *2.2. Complex relationship between urban form and resilience*

Many scholars have explored the reciprocal relationship between resilience and urban form. The existence of similarities between ecological systems and urban systems was highlighted [34]. They point out that the four constituents of urban systems are multi-scalar, networked, and often strongly coupled. In the similar sense, researchers have conceptualized the urban form as a multi-layered spatial-temporal system regulated by discrete cycles of growth, decay, and by the interplay of conservative top-down mechanisms and emergent bottom-up processes [35]. Moreover, ecological systems are also characterized by interdependent entities, relying on a sequence of historical events, exhibiting spatial connections, and possessing a nonlinear structure. Both systems exhibit considerable internal resilience within their respective domains of stability. These similarities reveal that the relation between form and resilience share the characteristics of nonlinearity, multifaceted, and intertwining [36].

As urban form is regarded as complex systems with many constituent components that exist at different scales, overlap and influence one another [17,25], each element, as an instrument through which resilient challenges respond, has profound repercussions on disturbances and shocks. They have their own adaptive cycles [35], behaviors and performance responsive to cues of changes [37]. Once combined as complex systems, either cannibalistic or synergistic impacts emerge from the relationship between all elements that constitute the system and with other interacting systems. For example, interplay between systems across scales shows that efforts on facilitating urban form resilience to flooding at city scale may cause burden at neighborhood scale. Compared to low-density areas, high-density neighborhoods often generate more local surface runoff because they have fewer green spaces and more built-up surfaces. However, the denser patterns seen in high-density settlements entail better capability to decrease total and global surface runoff at the citywide scale compared to the capability of low-density settlements due to the loss of green and open spaces in adjacent peri-urban areas are greater [38]. An additional observation of interaction between systems of the same scale highlighted those architectural structures and activities designed to prevent flooding, such as reservoir impoundment and water injections, could lead to heightened seismic risks. The injection of large volumes of fluid into subterranean formations for storage or disposal might cause some of the fluid to seep into fault lines, increasing pressure and triggering the release of stored energy. This process has the potential to induce seismic activity and, in some cases, earthquakes [39].

Moreover, the same features of urban form may cause mixed effects in terms of pandemic resilience. For example, cities with high population densities faced significant challenges in controlling the spread of the virus. However, these same dense urban areas often have better access to healthcare facilities, public transportation, and essential services, which can reduce mortality rates and enhance resilience in other ways. The extensive public transportation system enables essential workers to continue commuting, and the city's numerous parks and open spaces provide residents with areas to exercise and relax while maintaining social distancing [40]. Additionally, while much evidence supports the positive impact of green spaces on pandemic resilience, the urban green space strategies might be paradoxical in terms of environmental justice problems where the creation of new parks to facilitate neighborhood attraction and aesthetics can also increase housing cost and property value and exacerbate gentrification and polarity of society [41]. Resilience at one scale or to one disturbance may be achieved at the expense of other systems. These phenomena give prominence to potential resilience trade-off from the perspective of differences in spatial hierarchies, resilient objectives, and mixed effects.

### *2.3. Theoretical framework to link resilience and urban form*

The integration of resilience with urban form characteristics is not immediate. By tracing the theoretical development of the link between spatial morphology and urban resilience, the need becomes apparent for a unifying theoretical framework that can better integrate these fragmented areas of study. This recognition guides researchers toward adopting complex adaptive systems (CAS) theory, which primarily focuses on the macroscopic system structures and behaviors that result from the interactions between various subsystems at the microscopic level within complex systems. It offers robust tools for understanding the dynamics of change and interaction within and between urban systems over time. Such theoretical advancements are crucial for developing a more holistic understanding of how urban forms adapt and respond to challenges like pandemics. Through the vein of the CAS theory, researchers have explored the application of fundamentally ecological concepts and models to the urban form of cities [16,35], such as the Adaptive Cycle Model that describes the cyclical pattern of growth and decay of an ecological system, and the Panarchy theory that is characterized by discrete adaptive cycles of multiple parallel or cross-scale systems. Resilience, as an intrinsic trait ubiquitously represented through the dynamics of systems, waxes and wanes across separate phases and scales of cycles. Extended from both theories, its definition has evolved from Holling's measurable, descriptive concept to "a way of thinking" [9].

The Adaptive Cycle Model outlines four phases of system dynamics: Exploitation, conservation, release, and reorganization [34]. In the exploitation phase, resources are abundant, and resilience is high as the system rapidly grows and adapts to new opportunities. In cities, this mirrors periods of rapid development and infrastructure expansion. The conservation phase follows, where resources are accumulated, leading to rigidity and reduced adaptability. Cities in this phase focus on maintaining stability, but resilience decreases due to vulnerability to disturbances. The release phase occurs when accumulated resources are suddenly disrupted by events like economic downturns, disasters, or pandemics. Resilience is low as the system experiences significant change. In urban contexts, this could represent the breakdown of established structures, requiring cities to innovate and adapt to new challenges. Finally, the reorganization phase marks the rebuilding of resilience, fostering new structures and strategies. Cities may rethink urban planning to adapt to emerging realities and challenges. The Adaptive Cycle Model emphasizes that resilience is not static; it fluctuates over time as systems transition between phases. This dynamic nature reveals thresholds that dictate a system's ability to adapt and transform in response to disruptions [35].

Panarchy theory emphasizes that resilient systems consist of interconnected hierarchies, with faster changes occurring at lower levels and slower, more stable changes at higher levels [16]. Urban form, as a stable and spatial element, can support multiple processes simultaneously. For example, within an urban area, historic buildings may be preserved, maintaining heritage (a slow-changing process), while other structures are repurposed into modern uses like cafes or shops, attracting younger populations. Moreover, vacant lots may be redeveloped into apartments, and streets might undergo greening and pedestrianization efforts. These simultaneous developments demonstrate how urban form can accommodate various trajectories of change, preservation, adaptive reuse, new development, and environmental improvements, across different timelines and locations. This dynamic interplay of changes reflects the complex organizational structure of urban form, similar to systems observed in ecological contexts [34,35]. Urban form evolves through interconnected processes like historical succession and spatial interlocking, which occur non-linearly. This reinforces the notion that using resilience frameworks to analyze urban form in its different configurations is valid and beneficial, as it provides a comprehensive understanding of how cities can adapt to diverse challenges [18].

### *2.4. Quantitative assessment framework and urban resilience indicators*

The resilience theories outlined above provide the theoretical foundation and explanatory frameworks for understanding urban form resilience as nonlinear adaptive processes. To operationalize resilience in urban planning and design, many scholars point out the necessity to translate these abstract concepts into specific indicators that can be assessed and enhanced. Hillier et al. [42] bridge the gap from the view of space syntax where spatial elements are represented through geometric forms (nodes and edges). They point out the existence of duality with regard to urban street networks which is a central component of spatial form for the distribution of accessibilities. In the primal representation of spatial graphs, edges typically denote street segments, and nodes are the junctions where two or more edges intersect. In a dual representation, we can invert things, illustrating street segments as nodes and junctions as edges. The distinction between the two graphs lies in that the primal typically utilizes metric distances to describe interrelationships between graph elements, whereas the dual often focuses on topological distance measures and emphasizes the number of connections rather than the length of those connections between graph elements. This analysis method enables studies of relationships between spatial elements using measures such as integration (ease of access) and choice (passing flow), which help in understanding human movement and spatial accessibility. These methods make it possible to examine how spatial configurations are internalized into socio-economic processes.

Hillier et al. [42] dive deeper into the duality and point out that the "foreground network"

contributes to a part of duality that links city centers and primary movement with activity corridors across different scales of a city, maximizing connectivity and integration within the urban system. It comprises the physical arterial infrastructure that enables mobility, accessibility, and flow of resources during both normal temporal trajectories and disruptions. Another part of duality is manifested by the background residential network that provides spatial context shaped by local cultural factors and residential patterns. It forms the backdrop against which the foreground network operates and evolves and represents housing and neighborhoods that support the social fabric, community networks, and local resilience capacities. The two interconnected spatial networks shape urban form and experience, and the interplay between the two influences the resilience attributes [16]. For example, historically, London's development was significantly impacted by fires and aerial bombardments. After the Great Fire, there was no major redesign of the network or plot layout, but following World War II, despite some localized plot redesigns in the Comprehensive Development Areas and the construction of new roads like the M25, M4, and Westway, the basic radial structure of the existing main routes was preserved. Thus, the historical development of London can be seen because of a process of self-organization [43]. It also fits the Adaptive Cycle Model and the Panarchy theory by acknowledging interconnection of hierarchical structure and discrete life cycles of urban systems.

Feliciotti [18] and Marcus et al. [16] take a further step in trying to capture implications of urban form on resilience and their cascading impacts through a tripartite classification framework, system assets, system attributes, and system performances, which provides a systematic way to align spatial morphology with the principles of resilience thinking (Figure 2). System assets are defined as resources within an urban system necessary for city's operations, recovery, and adaptation. We focus on physical assets composed of common urban form elements. However, ambiguity and lack of precision exist when defining these crucial morphological components. Their definitions vary and depend on research questions at hand. Bobkova et al. [44] argue that these assets include plots, streets, and buildings. Other scholars extend the list to include blocks [18]. Dibble et al. [45] applied hierarchical clustering analysis with urban morphometrics and classified form components as plots, street-edges, blocks, streets, and sanctuary areas. They explain that each morphological scale frames lower-level components: "Sanctuary area" is the unit consisting of all lower-scale components, then streets frame blocks, blocks contain street edges, street edges are made up of series of plots bridging the transition from private to public spaces, and plots contain buildings or open areas, forming a compositional hierarchy. Researchers have also provided an evolutionary interpretation of urban change (i.e., urban form is constantly adapting to ever-changing, non-equilibrium conditions). For instance, they point out that, when a street edge is characterized with multiple plots, the more fine-grained the fabric is, the more conducive it is of informal interactions and experiential adaptations that the space encompasses. Thus, it shows higher persistence compared to those featuring only few large plots, ensuring that modularity is maintained over time, a precondition of structural complexity and hence efficiency [46].

As it is challenging to measure resilience directly, many studies resort to proxy variables, i.e., system attributes [16]. The literature does not converge on a unified list of attributes. For example, Ribeiro et al. [47] proposed redundancy, robustness, resourcefulness, and rapidity as reasonable indicators. Feliciotti et al. [48] identified five attributes frequently associated with urban form, i.e., diversity, connectivity, redundancy, modularity, and efficiency. Marcus et al. [16] pointed out that distance, diversity, and density are the fundamental variables of spatial form to understand urban resilience. These attributes are general in nature and are embedded in every facet of urban development, from the macro-scale of city-wide planning to the micro-scale of community-based initiatives. They are not specific to any disturbances, but influence the overall behavior of system. They interfere with system resilience performance via their distinct relationships with generic aspects of human use of space. For example, Marcus et al. [16] explained that, in principle, spatial distance correlates with connectivity and accessibility to human activity, spatial density with the amount of human activity, and spatial diversity with the differentiation of human activity. We have organized definitions and general resilience implications of commonly mentioned attributes as shown in Table 2. This enables the systematic discussion of these measures as measures of variations and enables us to conceptualize cities as a collection of physical components each possessing distinct spatial potentials for human activity.

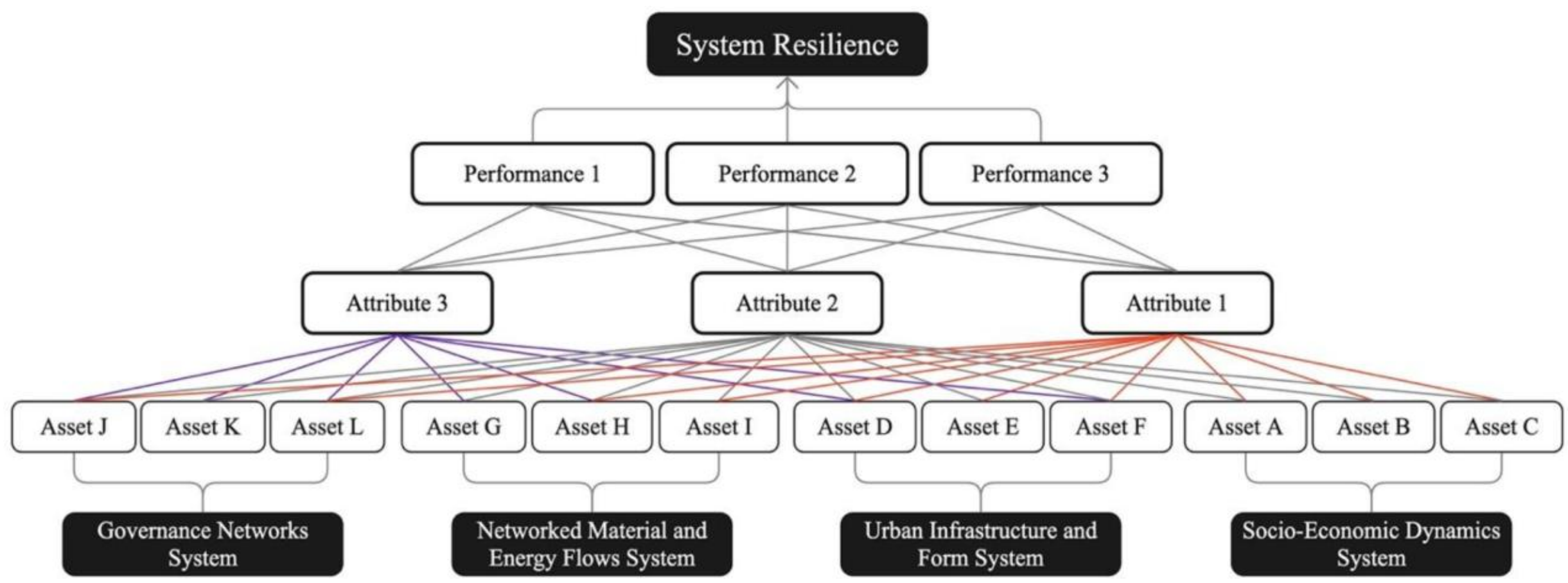

**Figure 2.** Conceptual framework of how resilience of urban systems is structured. It is a combinational effect from multiple system performances which are integrated results from several attributes as well. Moreover, interplays among constituent assets across subsystems determine characteristics of system attributes. Assets are grouped following the subsystem classification in Table 1. Note: Schematic drawing inspired by Dicken et al. [27] and Feliciotti [18].

There remains the heterogeneity of assessment techniques that lead to a paucity of systematic classification of metrics relating to resilience attributes. Some attributes might be interchangeable under certain contexts. For clarity, we update the resilience attributes assessment matrix (Table 3) initially organized by Feliciotti [18] and extend the measurements to six resilience attributes. These measurements have been widely referenced in the fields of urban design, sustainability science, and network science. We have also organized more detailed metric descriptions and measurement calculations in Table S1. It is recognized that the diversity of assessment metrics can introduce biases or inconsistencies when evaluating the same attributes. For example, differences in sensitivity to boundary definitions or the presence of outlier data can lead several measures of block and street connectivity to yield contradictory or skewed conclusions [18]. However, the proxy approach underscores an alternative to understand implications of urban form to resilience via separate measurable urban morphometrics. It would be prudent to consider a variety of measurements and perspectives to mitigate these weaknesses and establish a more reliable chain of evidence.

**Table 2.** Resilience attributes and corresponding spatial implications.

| System attributes | Definition and explanation | General resilience implication | System performance |
|---|---|---|---|
| Diversity | Recognized either functionally, as the presence of multiple distinct functions within proximity, or geometrically, as variation in the shape and size of urban components. In both scenarios, diversity correlates with more intensive space utilization and a greater capacity to support diverse uses and activities. | The fundamental attribute to support other attributes. Diverse urban forms create redundancy in urban systems, enable cities to adapt more easily to changing conditions, and ensures a more equitable distribution of resources and services, which can help communities better withstand and recover from disasters. | Adaptivity, Transformability, Recovery [18,49] |
| Connectivity | The degree to which various parts of a city are linked together through various networks. It encompasses the physical and functional connections that facilitate the movement of people, goods, and information within and between urban areas. Concepts associated with it include accessibility, interdependency, redundancy, integration, collaboration, inclusiveness and so on. | The relation between connectivity and resilience is complex. High connectivity can prevent isolation and speed up recovery by facilitating access to resources, but also enable disturbances to spread more quickly and make it difficult to isolate affected areas. While connections enable the exchange of information and resources, they can also reduce diversity and hinder local innovation and capacity building. | Adaptivity, Recovery [18,29,42] |
| Redundancy | Involves having multiple elements or systems that can perform the same function. In urban design, this can mean having multiple routes, services, or infrastructures that can take over if one fails. | Increases system robustness by providing backups and alternatives, ensuring that the urban form can continue to function even if some components are compromised. Additionally, the presence of multiple interchangeable components enables experimentation of novel solutions without risking the entire system, essential for cities aiming to transition to more sustainable urban forms. | Adaptivity, Transformability [9,18] |

| System attributes | Definition and explanation | General resilience implication | System performance |
|---|---|---|---|
| Modularity | Refers to the principle where a city or its components are divided into smaller, self-contained units or modules. These modules can function independently but can also be combined or reconfigured to form larger systems. Modularity is often associated with concepts such as independence or autonomy, self-sufficiency, flatness, compartmentalization, decentralization, and polycentricity. | Relative compartmentalization limits the extent to which shocks affecting a system preventing broader cascading failures. Fine grained urban fabrics, due to the challenges assembling smaller structures, tend to demonstrate greater persistence, maintaining modularity over time, crucial for structural complexity and efficiency. However, overly modular structures are prone to fragmentation, challenging coordination, and management. | Adaptivity, Transformability, Persistence [18,35,50] |
| Efficiency | Understood as the degree of structural complexity occurring at all scales of urban systems, as cities tend to enhance their structural efficiency by increasing organizational spatial complexity at every level. It also describes a specific organizational form of a system where relations between components and their assemblies are hierarchical, consistently observable at all levels in an ascending order of scales. Small, frequent components are balanced by fewer large, specialized ones. | The hierarchical structure allows a system to simultaneously execute multiple resilience strategies across different scales. It enables quick, small-scale adaptations at lower levels and maintains stability at higher levels, preventing rigidity. This structure also promotes cross-scale connections, encouraging synergies between components that can lead to transformative results. | Adaptivity, Transformability [16,18,48] |
| Density | Generally, refers to the concentration of people, dwelling units, bed units, habitual rooms, and activities within a given area. It can influence the efficiency of land use and the vibrancy of urban life by reflecting the intensity of population, built environment, employment, and activity. | Higher urban density enhances public service efficiency, supports extensive public transport, and reduces emissions by economizing on infrastructure and service delivery. It also helps preserve natural landscapes and manage environmental impacts through integrated green spaces. However, high-density areas are vulnerable to severe impacts, complicating recovery efforts. | Adaptivity, Transformability [16,18,36] |

**Table 3.** Resilience attributes assessment matrix. Note: Updated based on Feliciotti [18].

| | Sanctuary areas | Streets | Blocks | Street edges | Plots |
|---|---|---|---|---|---|
| Diversity | | | Block size | Street-edge interface matrix | Accessible plot density; Plot size heterogeneity |
| Connectivity | Avg. distance between urban main street | Link-node ratio; Avg. degree; Diameter; Permeability; Node density; Street density; Betweenness; Closeness; Straightness; Degree centrality | Block section; Block face length; Avg. degree; Permeability; Compactness; Closeness; Straightness; Degree centrality | | Avg. degree; Closeness; Straightness; Degree centrality; Gini-Simpson index |
| Redundancy | | Beta index; Meshedness coefficient; Cyclicity | | | Redundancy index |
| Modularity | Sanctuary area granularity | | | Street-edge granularity | |
| Efficiency | | Street length distribution; Network efficiency; Information centrality; Avg. Shimbel Index | | | |
| Density | Floor area ratio of Sanctuary area; Urban density; Population per total residential area; Built area density | Intersection density; Street length distribution | Dwelling units per block residential area; People per block residential area; People per block area | | Floor Area Ratio; Dwelling unit per plot area |

Although treating system attributes as proxies for assessment provides a handy toolkit to understand spatial implication of general resilience, there are gaps in the relationship between conceptual attributes and their measurement variables. It is not immediately apparent which metrics should be emphasized and what their specific resilience overtones are in the context of certain disturbances like large-scale public health emergencies. Hence, system performances serve as the essential role to pinpoint corresponding implications of urban form resilience. It represents the functional outcomes, capabilities, or behaviors of an urban system in response to specific changes [51,52]. Researchers have characterized resilience performance as an expression of system capacities representing what a specific system can or cannot do when confronted with a particular set of circumstances [18,53]. These researchers evaluate: i) The system's persistence or robustness in preserving the efficiency of its current state, which aligns with the definition of engineering resilience; ii) the system's adaptability and recovery, concentrating on its capability of maintaining critical functions, corresponding to the ecological resilience definition; and iii) the system's transformability or innovation capacity, referring to the ability to develop a fundamentally new system when ecological, economic, or social structures render the existing system unsustainable, in line with the evolutionary resilience definition [49,53]. The assessment involves quantifiable metrics [54] such as response time, service continuity level, the effectiveness of emergency response, and the rate of recovery of a system from a disturbance. This concept underscores the operational aspects of an urban system and enables urban planners and designers to understand the efficacy of an urban system's response mechanisms.

## 3. Discussion

### *3.1. Gaps in frameworks*

The frameworks discussed share similarities in their analysis methodologies, which involve considering system scales at varying resolutions. They alternate between coarse depictions of large-scale properties and fine-grained depictions for smaller-scale properties. The hierarchical abstraction of complex systems [55] portrays small-scale, fine-grained physical and concrete properties, corresponding to physical understandings of the elements of a system. While large-scale, coarse-grained properties are perceived as purposive and abstract, denoting the purposive understanding of why a system exists. This distinction reflects a common structure in human reasoning. The systematic mindset that advocates for a city's general views from urban form resilience to disruptions, as opposed to a specified perspective, aligns with the evolutionary notion that there is no equilibrium or so-called optimal state for urban systems but rather a prevalently ever-changing interplay between constituents [50,56]. It also prevents lopsided emphasis on one aspect while neglecting others; such methodologies implicitly adopt an event-and-situation-independent way to problem-solving across scenarios, which simplifies complex reasoning within a unified reasoning framework. Furthermore, these studies highlight that resilience should not be seen as a final state but rather as a characteristic of urban form that is sensitive to context. Its attributes may differ based on elements, such as the dynamics of time and space, the types of risks encountered, and the objectives sought [9,57]. If the study of urban form implications on pandemic resilience is to be placed within the context of general resilience, as suggested by the discussed frameworks, then a critical question emerges: Can the frameworks be directly applied to pandemic resilience?

COVID-19 has highlighted the need for a complex systems approach to investigate morphological implication of pandemic resilience in several ways [2,4]. First, knock-on effects of the virus, which span sectors, suggest that studies entailing only spatial factors may be narrow and inadequate to capture the full scope of interconnected, systemic challenges presented by the pandemic. However, the study of multifaceted interconnections may lead to the tradeoff between the breadth and the depth of analysis approaches. For example, school closures in New York City not only disrupted education, creating inequalities, but also impacted transportation and the economy, as many parents left the workforce or adjusted work routines [58,59]. The initial fear of the virus and shift to remote work spurred suburban housing demand, potentially signaling long-term urban migration and changing transportation patterns. According to panarchy theory [35], this represents a cross-scale mechanism of disruption, with fast-evolving viral impacts affecting the stable, slow-moving urban form. Each link in this chain of disturbances warrants further investigation to fully understand pandemic resilience in urban systems. This brings challenges in defining the boundary of pandemic impacts when studying the implication of urban form on pandemic resilience. Thus, the following questions arise: How broadly should we cast our net? At what point does our scope become so extensive that it obscures the underlying system dynamics we aim to understand?

Second, there is a key trade-off between the resolution of data and the geographic scale it covers. High-resolution data captures local dynamics and behaviors in detail, while low-resolution data provides a broader perspective over larger areas. However, inconsistencies in data collection across different regions can make comparisons difficult. For instance, GPS tracking can reveal individual movement patterns and virus transmission hotspots but may not be accessible everywhere due to technical, legal, or privacy issues, potentially missing broader trends [60–62]. Studies examining urban form's impact on pandemic resilience often use city-wide averages or proxies like changes in night-time light to measure socio-economic effects, though these approaches may miss localized variations. Moreover, the Modifiable Areal Unit Problem complicates spatial analysis, as conclusions drawn from aggregated data may not hold for individual cases, especially at smaller scales [63]. Although simplifications make findings more accessible for policymakers and the public, they might overlook critical localized nuances necessary for targeted interventions and strategies. While granular data offers depth, and coarse data provides breadth, neither approach is sufficient on its own.

Third, since the disruption of a pandemic will be felt for many years to come, there is a need to not only address immediate issues but longer-term consequences as well [2]. So far, gaps exist for analysis frameworks capable of addressing the matter at various time granularities. Research in urban pandemic resilience has primarily concentrated on comprehending the short-term tangible aspects of environmental quality, socio-economic consequences, management and governance, transportation, and urban design. Frequently, these researchers utilize index-based methodologies that select indicators according to categories such as economic stability, health systems, governance structures, and environmental conditions. Moreover, regression relations are established between chosen indicators and studied features. This type of method remains popular for evaluating resilience due to its practicality, multidimensional nature, and ability to enable easy horizontal comparisons [64,65]. However, the relative static frameworks are inadequate for capturing the dynamic changes and nonlinear interconnections within urban systems. Although a certain extent of temporal aspect can be included, such as constructing regression models for individual time windows, the causal relationship is short of interrogation. More specifically, there is limited knowledge regarding how short-term interactions could influence long-term city outcomes. Integrating short-term and long-term perspectives may be an overlooked research area that is crucial for achieving a resilient and sustainable urban environment.

Beyond these general observations, it is instructive to compare the mentioned frameworks with recent pandemic-specific studies that explicitly connect urban form to COVID-19 outcomes. Lak et al. [5] and AbouKorin et al. [6], for example, identified planning principles and urban form characteristics associated with pandemic resilience in cities ranging from Tehran to selected European metropolitan regions, while Banai [11], Sharifi [4], and Mouratidis [7] developed conceptual or normative frameworks around compactness, neighborhood design, and quality of life in the context of COVID-19. Venerandi et al. [8] showed how fine-grained morphometric descriptors of density, building form, and street-network structure can be related to spatial patterns of COVID-19 cases and deaths in Greater London. Together, these contributions confirm that the attributes we discuss, diversity, connectivity, modularity, redundancy, efficiency, and density, are empirically salient for pandemic resilience, but they are rarely organized within a unified, multi-scalar resilience framework.

Therefore, there is a need for a generalized framework that is adaptable and easily updated, enabling the straightforward integration of existing approaches within it. Flexibility enables more effective responses to diverse and changing conditions encountered during public health crises. Since the ways in which each event disrupts urban systems can differ, it is impossible to fully understand or predict every context and future scenario [66–68]. Also, as noted regarding the mixed effects of spatial features on the general resilience, empirical experience might not be directly applicable in the pandemic's context. Through an adaptive analysis framework with specific needs addressed, modeling specific problems within contexts can start with empirical data reflecting physical realities as the foundational basis. The cascading impacts of hazards can then be deduced by examining how these impacts interact and affect resilience properties. Resilience outcomes, whether short-term or long-term, can be discerned by examining the synergistic performances of various properties. It offers additional credibility to the choices and strategies for decision-makers in the face of emerging public health challenges.

### *3.2. Toward an adaptive framework*

Incorporating adaptability and generalization into the analytical framework might prompt the creation of a unified standardized system of indicators to assess urban resilience. Given that resilience varies with context, these indicators may not be universally applicable. A solution could be to generate quantifiable and location-specific results through bottom-up and top-down blend methods. For example, Bedinger et al. [68] modeled the systemic impacts of Covid in cities with the proposed Urban System Abstraction Hierarchy Model (USAH) to dissect the urban system into a five-level decomposition ranging from concrete to abstract elements. This provides a structured way to describe a system from its physical components to its overall functional purpose, helping to understand both how and why a system operates as it does. Applying the model at the city scale, USAH defines the lowest level of the model as City Resources (Level 5—Physical Objects), such as hospitals, which support specific Processes (Level 4—Object-Related Processes), like providing critical healthcare. These processes enable broader Tasks (Level 3—Generalized Functions), such as public health, leading to desired Outcomes (Level 2—Values & Priority Measures), like effective health safeguards. At the top are System Purposes (Level 1—Functional Purposes), such as social care. These system parts are nodes that are linked between levels through their functionality, connecting the physical to the abstract through a hierarchical network. Additionally, weights of links reflect adjacency status between nodes at distinct phases of a disruption. By tracking the importance changes of each node recorded by the corresponding eigenvector centrality, the change of resilience regime explained by the panarchy theory can be identified. The structure connects the elements by exploring why each node exists when moving up the hierarchy, and how it functions moving down, thereby linking the physical and abstract realms through their purposes and means.

However, due to the extensive range of involving elements and processes, the resilience analysis is considered aspatially; the approach is difficult to map resilience geographically. Cariolet et al. [56]

proposed the potential remedy to break down various aspects of resilience like working on a subset of resilience factors or focusing on the degraded mode of planned functioning of urban systems. Some researchers [64,65] narrowed their scope to subsystems of cities and take account of system dynamic methods to integrate the classic epidemiological dynamic model like the SIR model (i.e., Susceptible, Infectious, and Recovered) or the SEIR model (i.e., Susceptible, Exposed, Infectious, and Recovered) into the flow of analysis. Such an approach strongly couples the factor of time and can directly visualize the casual relationships, feedback loops, delay, and decision rules during the model simulation. Moreover, SD provides an open framework that can be linked with other disaster scenarios or probability simulation tools, making it suitable for calculating the resilience of urban systems under external shocks. Once the study unit is defined like block or census tract, it can be combined with spatial analysis to produce resilience maps and make sense of the dynamics of system elements and their interdependencies. In the future, researchers could incorporate more hierarchical reasoning into the workflow to make it easier and more straightforward for policy makers and stakeholders to understand how physical components operate alongside functional proposes of subsystems.

### *3.3. Scope and limitations*

This review has several limitations that should be acknowledged. First, it is intentionally designed as a narrative, critical synthesis of conceptual and methodological frameworks rather than as a fully systematic or bibliometric review. As a result, we do not claim to have exhaustively captured all relevant publications on urban resilience, urban morphology, or pandemic resilience, and our selection of studies inevitably reflects judgement about their conceptual influence and relevance. Second, our corpus is largely limited to English-language peer-reviewed literature, which may under-represent work published in other languages and in policy or practitioner outlets. Third, while we draw on pandemic-specific studies, we do not provide a comprehensive empirical comparative analysis of COVID-19 outcomes across cities; instead, we concentrate on how general resilience and morphometric frameworks can inform such analyses. These limitations also point to future research needs, including systematic mapping of pandemic-resilience indicators across a wider range of contexts and empirical testing of the relationships implied by our conceptual framework.

## 4. Conclusion

Through the lens of theoretical and methodological approaches, we have identified key insights and gaps that underscore the need for more adaptive and systematic frameworks for assessing urban resilience. Urban form significantly influences a city’s resilience to disruptions, including pandemics. Urban form’s design and structure can either enhance or hinder a city’s ability to manage and recover from such events. This influence extends to the movement of people, goods, and services, thereby playing a vital role in pandemic preparedness and response. Several methods and models, such as system dynamics, complex adaptive systems (CAS) theory, and assessments of resilience attributes, have been utilized to measure urban resilience. Nonetheless, the diversity of assessment methods creates obstacles in forming a standardized approach.

Although there have been advancements in understanding how urban resilience and form interact, considerable gaps and opportunities for enhancement persist. The concept of resilience must be dynamic, context-sensitive, and adaptable to diverse types of disruptions, not just pandemics.

Moreover, models often lack the necessary flexibility or do not sufficiently consider the dynamic nature of urban environments. Thus, there is a need for frameworks that are responsive to evolving circumstances and can incorporate new data and insights on an ongoing basis. Additionally, many researchers concentrate on the immediate effects and short-term resilience strategies, overlooking the long-term sustainability of urban systems. In the future, researchers should integrate immediate and extended strategies to craft thorough resilience plans that cater to current and future requirements.

This review contributes to the ongoing discourse on urban resilience by highlighting the critical role of urban form in pandemic resilience and broader urban resilience frameworks. Moving forward, the development of more adaptive and comprehensive frameworks is crucial. These frameworks should be used to integrate multi-dimensional data, consider various scales and temporal aspects, and be adaptable to varied contexts and disruptions. Such advancements will better equip cities to withstand and recover from forthcoming crises.

## Use of AI tools declaration

The authors declare they have not used Artificial Intelligence (AI) tools in the creation of this article.

## Acknowledgments

The research is based upon work supported by National Science Foundation under Award No. 2033580. Any opinions, findings, and conclusions or recommendations expressed in this material are those of the authors and do not necessarily reflect the views of the National Science Foundation.

## Conflict of interest

The authors declare that there are no conflicts of interest regarding the publication of this paper.

## Author contributions

Y.S., R.W., A.S., T.D., and S.S. designed research; Y.S. organized data and literature; Y.S. and R.W. performed research; Y.S. and R.W. wrote the paper; Y.S., R.W. and A.S. modified the paper.

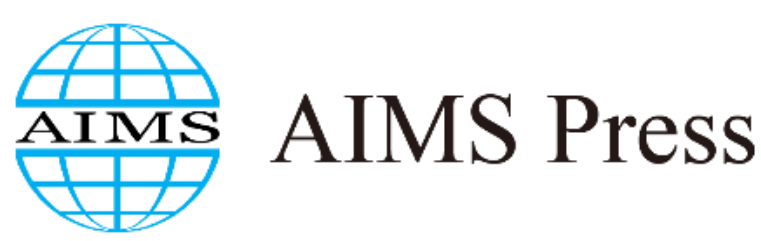